\documentclass[preprint,12pt,authoryear]{elsarticle}

\usepackage{amssymb}
\usepackage{amsmath}

\usepackage{lineno}

\usepackage{verbatim}

\journal{Meteoritics and Planetary Science}

\begin{document}

\begin{frontmatter}



\title{Chondrule size averaging in two and three dimensions} 


\author[MNHN]{Emmanuel Jacquet} 


\affiliation[MNHN]{organization={Institut de Min\'{e}ralogie, de Physique des Mat\'{e}riaux et de Cosmochimie, UMR7590, Mus\'{e}um national d'Histoire naturelle, CNRS},
            addressline={CP52, 57 rue Cuvier}, 
            city={Paris},
            postcode={75005}, 
            country={France}}

\begin{abstract}
Chondrule sizes are an important discriminant of chondrite groups and chondrule formation theories. Yet most studies report 2D sizes without attempting the somewhat tedious corrections to 3D proposed in the past literature. I show here that 3D size averages can be simply calculated directly from 2D averages of appropriate powers of sizes (and vice-versa), without having to reconstruct 3D distributions, although explicit formulas to that end are also provided. This illuminates the comparison between 3D and 2D arithmetic size averages, which are confirmed to be little different for the published datasets. Yet I suggest another type of average, the "sphere-weighted average" (i.e. weighted by the chondrule equal-volume sphere area), a measure of the surface/volume ratio of the chondrule population, which is more robust against omission of small chondrules. While larger than the arithmetic averages, it is best in line with previous literature estimates based on less exhaustive surveys of chondrules.
\end{abstract}

\end{frontmatter}



\section{Introduction}

  Chondrule are (sub)millimeter-size igneous spheroids ubiquitous in primitive meteorites, but whose origin remains elusive. While most of the attention has been devoted to their thermal history \citep[e.g.][]{Jonesetal2018} or ambient solid densities \citep[e.g.][]{Alexanderetal2008,CuzziAlexander2006,Tenneretal2015,Jacquetetal2026}, their sizes can also inform chondrule (precursor) formation and/or sorting processes, be it by their average value \citep[e.g.][]{Dodd1976,Jacquet2014size} or the shapes of their probability distribution \citep[e.g.][]{Cuzzietal2001,Teitleretal2010}. They are also useful classification diagnostics, e.g. between ordinary chondrite groups \citep{Jones2012}. A comprehensive compilation until 2015 can be found in \citet{Friedrichetal2015}.

  While some size distribution data have been obtained by disaggregation \citep[e.g.][]{MartinMills1978,Metzler2018} or, more recently, computed tomography \citep{Friedrichetal2022,Floydetal2024}, most pertain to 2D thin or thick sections of chondrites \citep[e.g.][]{Dodd1976,RubinGrossman1987,Rubin1989,NelsonRubin2002,Ebeletal2016,Ebeletal2024,FendrichEbel2021,Simonetal2018,Metzler2018,Friendetal2018,Schneideretal2002,Charlier2025}.  However, 2D size distributions certainly differ from 3D ones. Indeed, chondrules are generally not sectioned in an equatorial plane, such that their 2D size underestimates their 3D size, and the sectioning probability is greater for bigger chondrules, leading to an opposite effect \citep{Eisenhour1996}. Assuming spherical shapes for chondrules, it is relatively straightforward (if somewhat tedious) to express the 2D probability distribution functions (PDF) as a function of the 3D PDF, but the inversion from 2D to 3D has been hitherto performed only numerically \citep{Eisenhour1996,CuzziOlson2017}, with no direct estimate of the error made e.g. as to the average size. As a result, most publications reporting 2D data do not attempt any correction to 3D \citep{Friedrichetal2015}.

  In this work, I show that 3D size averages can be very simply extracted directly from 2D data, with an estimate of the statistical error, without having to calculate a 3D PDF. And this even though the simple proof of this can be used to produce an explicit expression of that 3D PDF, which I also provide. After the general demonstrations (which can be skipped by the practical reader), I discuss the comparison between 2D and 3D arithmetic size averages as a function of the relative standard deviation of the 3D size, and propose a new "sphere-weighted" size average, a function of the overall surface/volume ratio which is more robust against omission of the smallest chondrules.

\section{General proof}

  In the following I denote by $a_{3D}$ the 3D radius of a chondrule (assumed to be spherical) and $a_{2D}$ that of a particular 2D section thereof (which depends on how close the plane of section passes to its 3D center). I ignore the effect of a finite thickness of a thin section examined in transmission \citep[e.g.][]{Eisenhour1996}, with most recent studies being now performed on back-scattered electron images. 

\begin{figure}
\centering
\includegraphics[width=\textwidth]{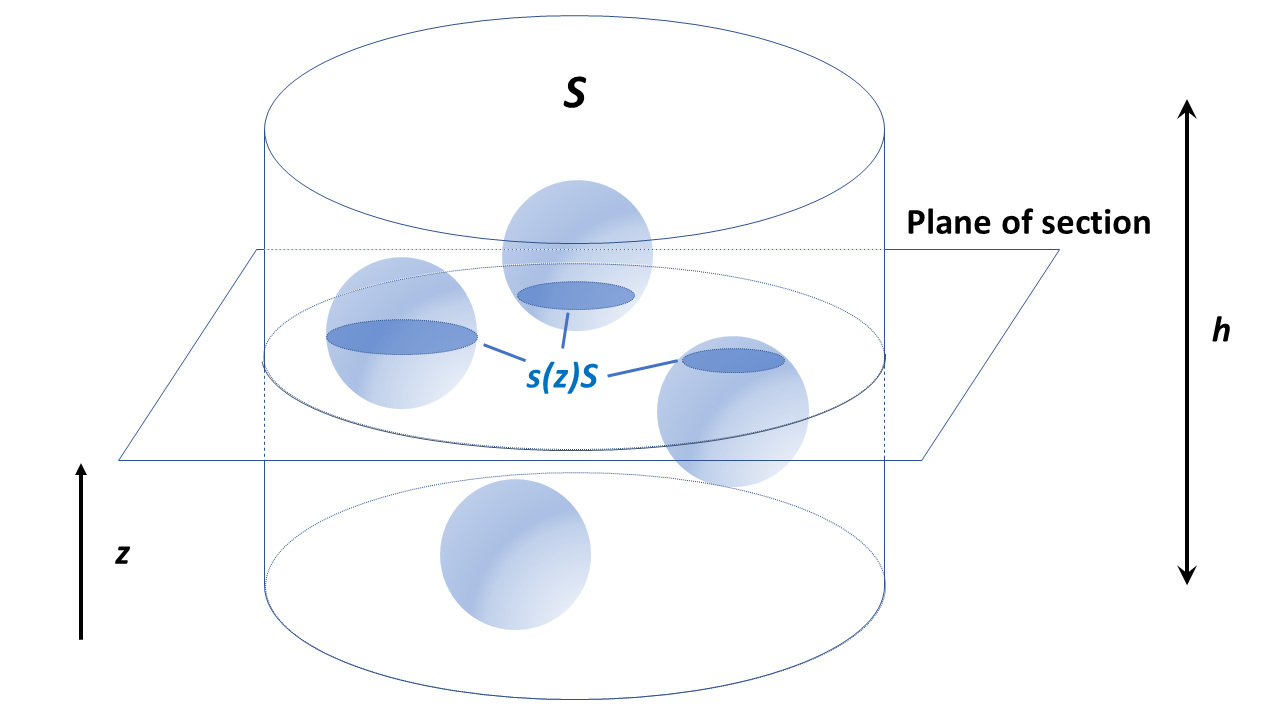}
\caption{Sketch of the statistical equivalence between volume and surface fraction in a cylinder. The component of interest consists in (four) blue chondrules. Their intersections with the plane of section (inside the cylinder) at height $z$ make a total area $s(z)S$ with $S$ the base of the cylinder. 
}\label{cylinder}
\end{figure}

  We first recall that for any component, the expectation value of its volume fraction $v$ is the same as that of its 2D surface fraction $s$. Indeed, let us imagine a cylinder of base $S$ comprised between $z=0$ and $z=h$, carved out a statistically uniform volume (Fig. \ref{cylinder}). Then:
\begin{equation}
v=\frac{1}{Sh}\int_0^{h}s(z)S\mathrm{d}z = \frac{1}{h}\int_0^{h}s(z)\mathrm{d}z 
\end{equation} 
that is, the average of $s$ between $z=0$ and $z=h$. Thus the expectation values (averaging over all statistical fluctuations) must coincide.

  We are now interested in expressing the area-weighted average of any function $f(a_{2D})$, which we denote by $\langle f(a_{2D})\rangle_{\pi a_{2D}^2}$ (where the subscript indicates the weighing function), as a function of the 3D PDF. It will be then a matter of choosing the right $f$ to conversely express 3D averages of interest as a function of 2D averages.

\begin{figure}
\centering
\includegraphics[width=\textwidth]{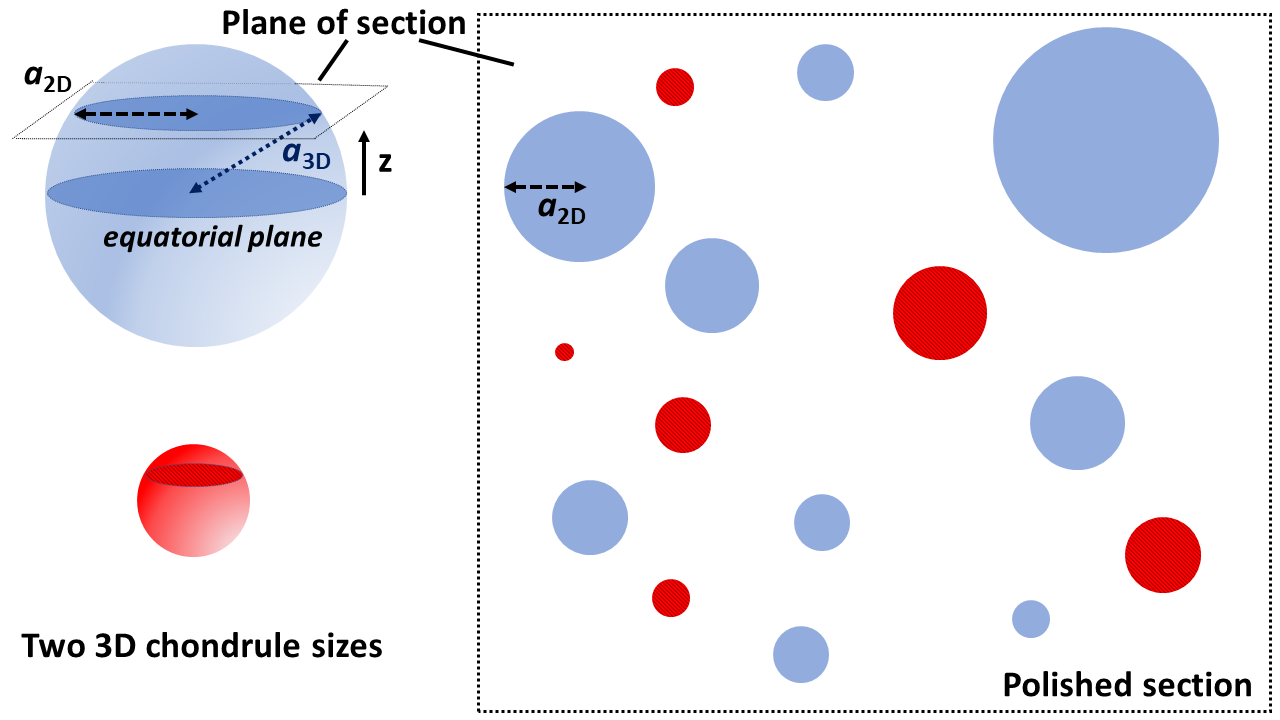}
\caption{Illustration of the basic argument in the case of a size distribution restricted to two 3D sizes (with the red chondrules being smaller than the blue chondrules). Depending on the offset $z$ between the plane of section and the parallel equatorial plane, the 2D radius $a_{2D}$ will be smaller than the 3D radius $a_{3D}$, hence the various 2D sizes of the blue 2D cross sections (same for the red chondrules). 
}\label{sketch}
\end{figure}

  We first envision the case of a distribution with two (3D) sizes only, with the blue chondrules having a 3D size $a_{\rm blue}$ bigger than the red chondrules' ($a_{\rm red}$; Fig. \ref{sketch}). When calculating the area-weighted average of some function $f(a_{2D})$ over all chondrule sections in the plane, we can first average each color (i.e. 3D size) separately. Then the results for each 3D size (which we can denote by $\langle f(a_{2D})\rangle_{\pi a_{2D}^2;a}$ with $a$ the fixed 3D size) can be averaged weighted by the areas of all their 2D sections. These correspond to the 3D volume proportions of the two populations following the above argument. If the volume fraction (among chondrules) of the blue chondrules is $v_{\rm blue}$, we thus have:
\begin{equation}
\langle f(a_{2D})\rangle_{\pi a_{2D}^2}=v_{\rm blue}\langle f(a_{2D})\rangle_{\pi a_{2D}^2;a_{\rm blue}}+(1-v_{\rm blue})\langle f(a_{2D})\rangle_{\pi a_{2D}^2;a_{\rm red}}
\end{equation}

  This can be generalized to any size distribution as:
\begin{equation}
\label{f(a2D) original}
\langle f(a_{2D})\rangle_{\pi a_{2D}^2}=\langle\langle f(a_{2D})\rangle_{\pi a_{2D}^2;a_{3D}}\rangle_{4\pi a_{3D}^3/3}
\end{equation}

That is, as the inconsequential factors $\pi$ or $4\pi/3$ can be dropped:
\begin{equation}
\label{f(a2D)}
\frac{\langle a_{2D}^2 f(a_{2D})\rangle}{\langle a_{2D}^2\rangle}=\frac{\langle a_{3D}^3\langle f(a_{2D})\rangle_{a_{2D}^2;a_{3D}}\rangle}{\langle a_{3D}^3\rangle},
\end{equation}
with the lack of subscript denoting an arithmetic average.

  For a given $a_ {3D}$, the offset $z$ of the plane of section with respect to the parallel equatorial section (Fig. \ref{sketch}) is uniformly distributed between -$a_{3D}$ and $a_{3D}$. Since $a_{2D}^2=a_{3D}^2-z^2$:
\begin{eqnarray}
\label{f(a2D) at fixed a3D}
\langle f(a_{2D})\rangle_{a_{2D}^2; a}&=&\int_{-a}^{a}f\left(\left(a^2-z^2\right)^{1/2}\right)\left(a^2-z^2\right)\mathrm{d}z   \left(\int_{-a}^{a}\left(a^2-z^2\right)\mathrm{d}z\right)^{-1}\nonumber\\
&=& \frac{3}{2}\int_0^1 f\left(a(1-Z^2)^{1/2}\right)\left(1-Z^2\right)\mathrm{d}Z\nonumber\\
&=& \frac{3}{2}\int_0^{\pi/2} f\left(a\mathrm{sin}\theta\right)\mathrm{sin}^3\theta\mathrm{d}\theta
\end{eqnarray}

  In particular, for a constant exponent $q$:
\begin{equation}
\label{a2D for fixed a3D}
\langle a_{2D}^q\rangle_{a_{2D}^2; a}=\frac{3}{2}a^q W_{q+3}
\end{equation}
with the Wallis integral
\begin{equation}
W_n\equiv\int_0^{\pi/2}\mathrm{sin}^n\theta\mathrm{d}\theta
\end{equation}
tabulated in table \ref{Wn table}. \ref{Ellipsoids} examines how (weakly) the result is modified for ellipsoidal chondrules.

\begin{table}
\caption{Values of the Wallis integral. The two final columns are meant for an integer $p$.}
\label{Wn table}
\begin{tabular}{c c c c c c c c}
\hline
$n$ & 0 & 1 & 2 & 3 & 4 & $2p$ & $2p+1$\\
\hline
$W_n$ & $\frac{\pi}{2}$ & 1 & $\frac{\pi}{4}$ & $\frac{2}{3}$ & $\frac{3\pi}{16}$ & $\frac{(2p)!\pi}{2^{2p+1}p!^2}$ & $\frac{2^{2p}p!^2}{(2p+1)!}$ \\
\hline
\end{tabular}
\end{table}

We thus obtain:
\begin{equation}
\label{a2D-3D^q weighted}
\frac{\langle a_{2D}^{q+2}\rangle}{\langle a_{2D}^2\rangle}=\frac{3W_{q+3}}{2}\frac{\langle a_{3D}^{q+3}\rangle}{\langle a_{3D}^3\rangle}
\end{equation}

Dividing this equation for $q=p-2$ by the same for $q=-2$ yields:

\begin{equation}
\label{a2D^p}
\langle a_{2D}^p\rangle = W_{p+1}\frac{\langle a_{3D}^{p+1}\rangle}{\langle a_{3D}\rangle}
\end{equation}

Conversely, dividing equation \ref{a2D-3D^q weighted} for $q=p-3$ by the same for $q=-3$ yields:
\begin{equation}
\label{a3D^p}
\langle a_{3D}^p\rangle = \frac{\pi}{2W_p} \frac{\langle a_{2D}^{p-1}\rangle}{\langle a_{2D}^{-1}\rangle}
\end{equation}

The relationships derived herein thus do not require expressing the 3D PDF, nor did their proof. Nevertheless, \ref{Inversion} extracts it as a corollary and even shows that our basic theorem (equation \ref{f(a2D) original}) can be rederived, if more tediously, from a more traditional expression of the "forward problem" (from 3D to 2D).

While all above equations have been derived for chondrule \textit{radii} $a$, the very same ones apply when \textit{diameters} are meant instead.

\section{Application}

\subsection{Arithmetic average}
 
Equation \ref{a3D^p} evaluated for $p=1$ yields a simple expression of the arithmetic 3D size ($a_{3D}$) average proportional to the harmonic average of the 2D size ($a_{2D}$):
\begin{equation}
\label{a3D}
\langle a_{3D}\rangle = \frac{\pi}{2\langle a_{2D}^{-1}\rangle}
\end{equation}
So assuming $n$ 2D measurements $a_{2D,1}, ... a_{2D,n}$, a good estimator\footnote{See \ref{Error} for a slight $O(1/n)$ bias correction.} of $\langle a_{3D}\rangle$ would be $\pi n/(2\sum_{i=1}^{n}a_{2D,i}^{-1})$.

  Similarly, equation \ref{a2D^p} yields:
\begin{equation}
\langle a_{2D}\rangle = \frac{\pi\langle a_{3D}^2\rangle}{4\langle a_{3D}\rangle}
\end{equation}

\begin{figure}
\centering
\includegraphics[width=\textwidth]{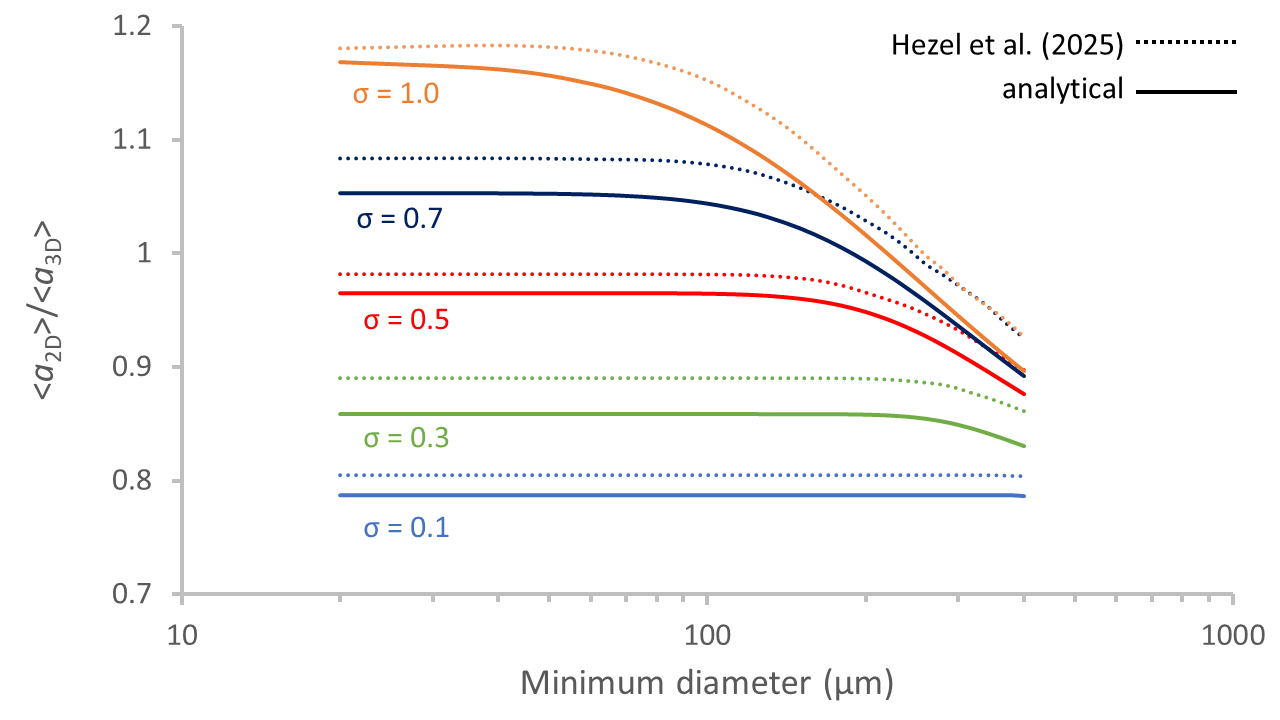}
\caption{Comparison of the ratio of 2D and 3D arithmetic averages in the case of a log-normal distribution with (log) mean $\mu=6.2$ (for sizes expressed in microns) truncated at 1500 $\mu$m and a minimum threshold taken as the abscissa \citep{Hezeletal2025}. Curves are drawn for different value of the (log) standard deviation $\sigma$, and decrease since higher minima make the distribution narrower (hence a lower relative standard deviation, see equation \ref{a2D and RSD(a3D)}). The values obtained in the simulations of \citet{Hezeletal2025}, plotted in dotted lines, are slightly offset because they estimated the 2D average via a log-normal fit of their output 2D size distribution rather than directly averaging it.}\label{Hezel}
\end{figure}

  This illuminates at once the comparison of the 2D and 3D averages. Indeed:
\begin{equation}
\label{a2D and RSD(a3D)}
\frac{\langle a_{2D}\rangle}{\langle a_{3D}\rangle}=\frac{\pi}{4}\left(1+RSD(a_{3D})^2\right)
\end{equation}
with "RSD" denoting the relative standard deviation. Essentially, the $\pi/4$ factor (alone for a monodisperse chondrule population) reflects the effect of nonequatorial sectioning of a given chondrule \citep[e.g.][]{Hughes1978} while the $1+RSD(a_{3D})^2$ factor stems from the overrepresentation of larger chondrules among sectioned objects \citep[e.g.][]{Eisenhour1996}. The 2D average surpasses the 3D average if and only if $RSD(a_{3D})>\sqrt{4/\pi-1}\approx 0.53$. 
For the \citet{Hezeletal2025} example of a log-normal distribution truncated at a minimum  $a_{\rm min}$ and a maximum $a_{\rm max}$, we have:
\begin{eqnarray}
\frac{\langle a_{2D}\rangle}{\langle a_{3D}\rangle}=\frac{\pi}{4}e^{\sigma^2}\left(\mathrm{erf}\left(\frac{\mathrm{ln}a_{\rm max}-\mu}{\sigma\sqrt{2}}-\sqrt{2}\sigma\right)-\mathrm{erf}\left(\frac{\mathrm{ln}a_{\rm min}-\mu}{\sigma\sqrt{2}}-\sqrt{2}\sigma\right)\right)\nonumber\\
\left(\mathrm{erf}\left(\frac{\mathrm{ln}a_{\rm max}-\mu}{\sigma\sqrt{2}}\right)-\mathrm{erf}\left(\frac{\mathrm{ln}a_{\rm min}-\mu}{\sigma\sqrt{2}}\right)\right)\nonumber\\
\left(\mathrm{erf}\left(\frac{\mathrm{ln}a_{\rm max}-\mu}{\sigma\sqrt{2}}-\frac{\sigma}{\sqrt{2}}\right)-\mathrm{erf}\left(\frac{\mathrm{ln}a_{\rm min}-\mu}{\sigma\sqrt{2}}-\frac{\sigma}{\sqrt{2}}\right)\right)^{-2}
\end{eqnarray}
with $\mu$ and $\sigma$ the (log) mean and standard deviation of the untruncated distribution. This is plotted in Fig. \ref{Hezel}. If truncations can be ignored, the 2D average surpasses the 3D average for $\sigma > \sqrt{\mathrm{ln}(4/\pi)}\approx 0.49$.

\begin{figure}
\centering
\includegraphics[width=\textwidth]{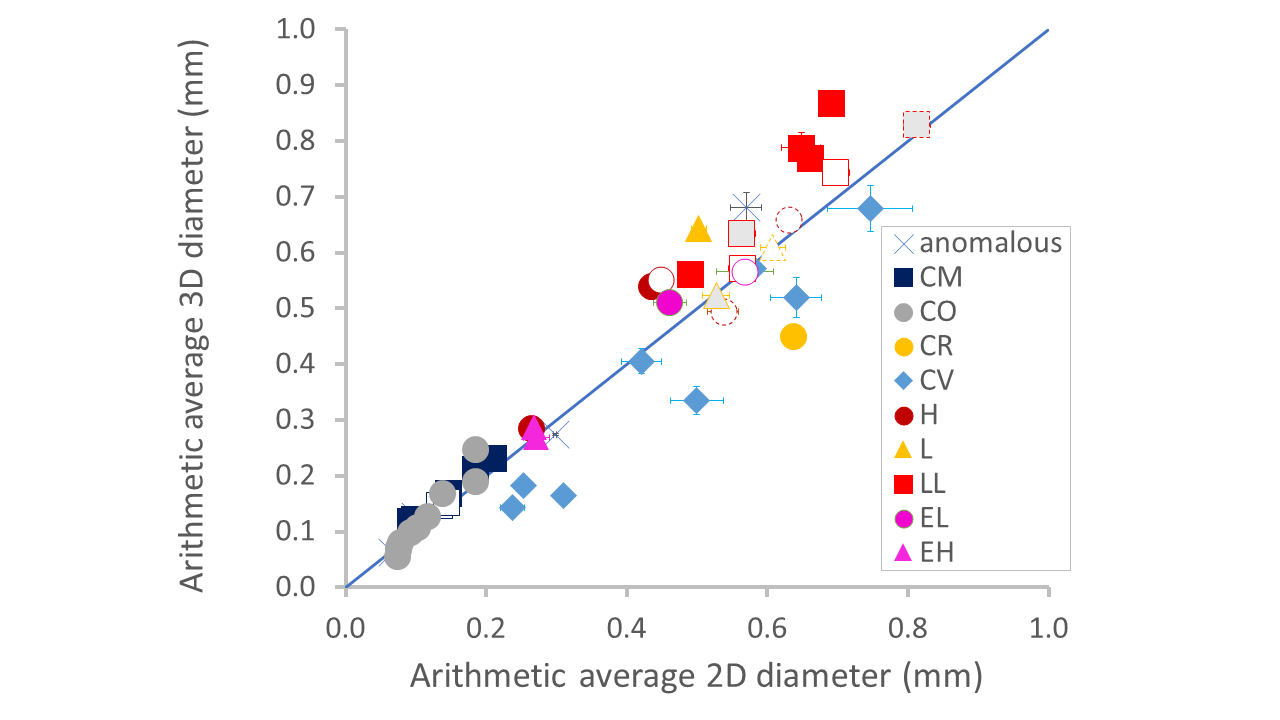}
\caption{3D vs. 2D size averages. Closed symbols represent 2D data and open symbols 3D data (with, respectively, 3D and 2D moments evaluated thanks to this work), with those with dashed outlines from disaggregation studies (the others being from computed tomography (CT)). Error bars are one standard deviation (cf \ref{Error}). The 1:1 line (i.e. abscissa = ordinate) is also drawn for reference. Data sources tabulated in table \ref{size table}. 
}\label{2D vs 3D}
\end{figure}

\begin{figure}
\centering
\includegraphics[width=\textwidth]{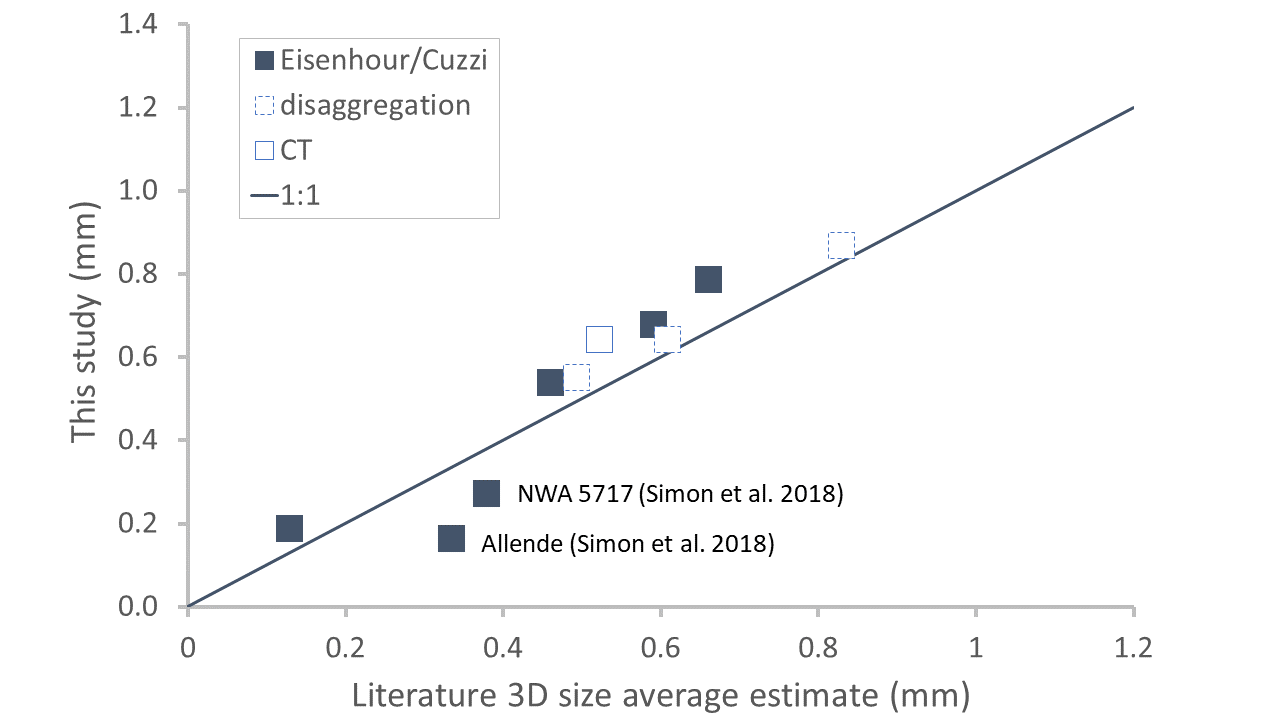}
\caption{Comparison of our 3D average diameters with previous estimates based on the \citet{Eisenhour1996}/\citet{CuzziOlson2017} method, disaggregation studies \citep{Metzler2018} or CT \citep{Friedrichetal2022} for matched meteorites.}\label{3D vs litt}
\end{figure}

\begin{figure}
\centering
\includegraphics[width=\textwidth]{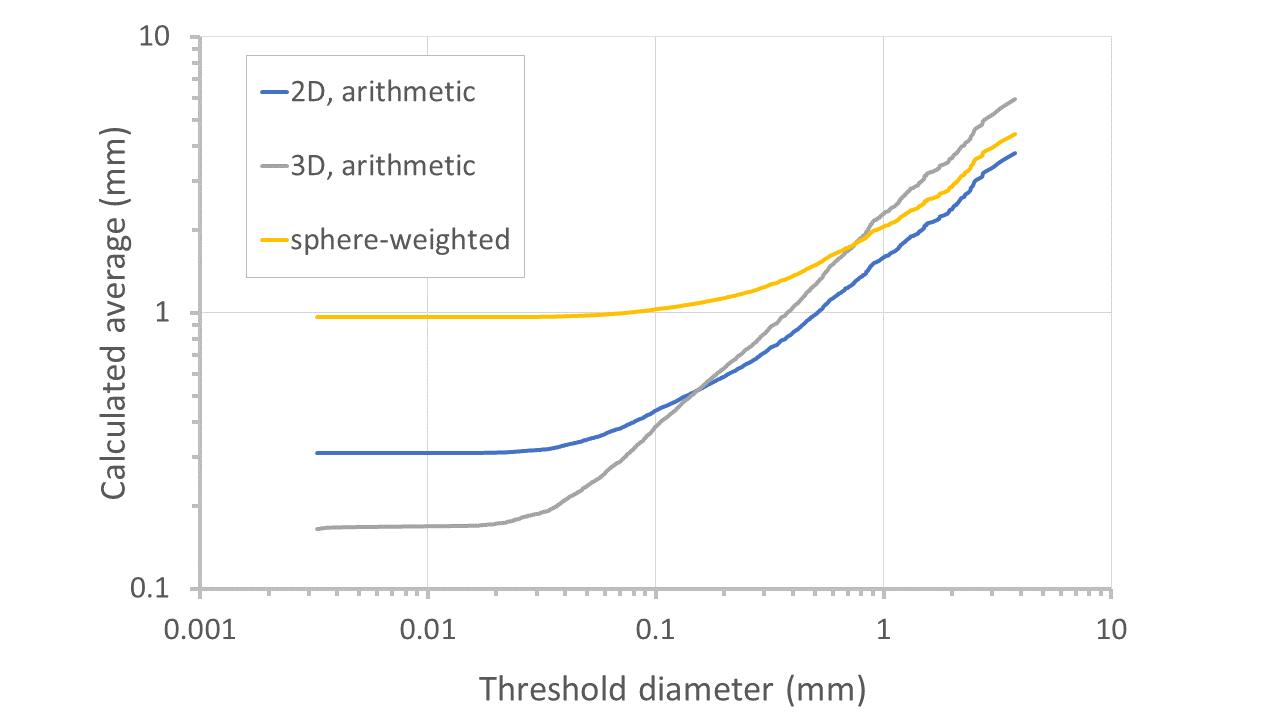}
\caption{Dependence of Allende chondrule size averages depending on the minimum size selected, in the \citet{Simonetal2018} dataset. The sphere-weighted average is more robust than its arithmetic counterparts, and hardly changes for thresholds below 100 $\mu$m.}\label{Allende threshold}
\end{figure}

On the whole, 2D and 3D average of actual chondrule distribution match within $\sim$20 \% (relative standard deviation of the ratio in our dataset; Fig. \ref{2D vs 3D}). The values of these averages are compiled in table \ref{size table}.  In general, our 3D averages fairly match with previous inversions or disaggregation studies (Fig. \ref{3D vs litt}), although the error associated with the former are not known. Nevertheless, our Allende size average of 0.165$\pm$0.004 mm is half that reported by \citet{Simonetal2018}, who reported an "unfolded" mean of 0.334 mm and a standard deviation of 0.458 mm. This is not reliable as such a high relative standard deviation would imply a $\langle a_{2D}\rangle=0.76$ mm following equation  \ref{a2D and RSD(a3D)}, more than twice the observed value of 0.33 mm. In fact, we suggest that even our estimate is not robust because it is sensitive to the threshold chosen for small chondrules (Fig. \ref{Allende threshold}). This is a general property of arithmetic averaging. The better resolution and systematicity of recent studies have led to averages significantly lower than previous literature simply because more small chondrules were included \citep[e.g.][]{Simonetal2018,Ebeletal2024}, but these are increasingly epiphenomenal as to the bulk properties of the chondrule population (total volume, surface...).  This prompts us to suggest an alternative metric in the following subsection.


\begin{table}
\caption{Chondrule diameter average (mm). Technique : D = disaggregation, PTS = polished thin/thick section, CT = computed tomography. Italicizing denotes that the published datasets were binned; each bin was ascribed to its mean size (geometric for logarithmic bins) to calculate all averages for consistency, but those actual 3D arithmetic averages of unbinned data reported by the authors are 0.59 mm and 0.57 mm for Bjurb\"ole and Allegan, respectively. Errors are one standard error, calculated following appendix C. Subtypes adapted after \citet{JacquetDoisneau2024}.}
\label{size table}
\resizebox{\textwidth}{!}{\begin{tabular}{c c c c c c c c}
\hline
Meteorite & Classification & Reference & Technique & 2D arithmetic & 3D arithmetic & sphere-weighted & $n$\\
\hline
Acfer 094 & C3.0-an & \citet{Ebeletal2016} & PTS & 0.067$\pm$0.002 & 0.062$\pm$0.002 & 0.144$\pm$0.010 & 598  \\
Asuka 12236 & C2.8-an & \citet{Charlier2025} & PTS & 0.100$\pm$0.002 & 0.128$\pm$0.001 & 0.175$\pm$0.009 & 2126\\
\hline
Jbilet Winselwan & CM2.3 & \citet{Friendetal2018} & PTS & 0.146$\pm$0.005 & 0.168$\pm$0.004 & 0.265$\pm$0.018 & 508\\
LEW 85311 & CM2.4 & \citet{Floydetal2024} & CT & 0.134$\pm$0.006 & 0.147$\pm$0.005 & 0.202$\pm$0.016 & 154\\
Murchison & CM2.3 & \citet{FendrichEbel2021} & PTS & 0.184$\pm$0.016 & 0.213$\pm$0.009 & 0.407$\pm$0.130 & 118\\
Murchison & CM2.3 & \citet{Charlier2025} & PTS & 0.092$\pm$0.002 & 0.118$\pm$0.001 & 0.161$\pm$0.008 & 1369\\
Paris & CM2.5 & \citet{Charlier2025} & PTS & 0.097$\pm$0.001 & 0.123$\pm$0.001 & 0.177$\pm$0.006 & 7758\\
Winchcombe & CM2.1 & \citet{Floydetal2024} & CT & 0.144$\pm$0.005 & 0.153$\pm$0.005 & 0.217$\pm$0.009 & 204\\
\textit{various} & CM2 & \citet{Floydetal2024} & PTS & 0.209$\pm$0.005 & 0.232$\pm$0.005 & 0.352$\pm$0.012 & 763\\
\hline
Colony & CO3.0 & \citet{Ebeletal2016} & PTS & 0.074$\pm$0.001 & 0.056$\pm$0.001 & 0.162$\pm$0.008 & 2247\\
El M\'edano 216 & CO3.2 & \citet{Pintoetal2021} & PTS & 0.092$\pm$0.002 & 0.100$\pm$0.002 & 0.168$\pm$0.008 & 1231\\
El M\'edano 463 & CO3.2 & \citet{Pintoetal2021} & PTS & 0.116$\pm$0.002 & 0.128$\pm$0.001 & 0.218$\pm$0.011 & 2792\\
Isna & CO3.8 & \citet{Pintoetal2021} & PTS & 0.184$\pm$0.004 & 0.248$\pm$0.005 & 0.256$\pm$0.008 & 343\\
Kainsaz & CO3.2 & \citet{Ebeletal2016} & PTS & 0.075$\pm$0.002 & 0.068$\pm$0.001 & 0.151$\pm$0.007 & 1625\\
Lanc\'e & CO3.2 & \citet{Ebeletal2016} & PTS & 0.077$\pm$0.002 & 0.081$\pm$0.002 & 0.140$\pm$0.005 & 1330\\
Los Vientos 123 & CO3.0 & \citet{Pintoetal2021} & PTS & 0.075$\pm$0.001 & 0.073$\pm$0.001 & 0.149$\pm$0.006	& 2319\\
Ornans & CO3.2 & \citet{Ebeletal2016} &	PTS	& 0.102$\pm$0.003	& 0.109$\pm$0.003 &	0.184$\pm$0.011 & 486\\
Warrenton & CO3.7	& \citet{Ebeletal2016} &	 PTS &	0.138$\pm$0.004 & 0.168$\pm$0.004	& 0.215$\pm$0.011	& 427\\
\textit{various} &	CO3	& \citet{Rubin1989}	& \textit{PTS}	& 0.184$\pm$0.003	& 0.190$\pm$0.003	& 0.353$\pm$0.018	& 2834\\
\hline
Renazzo	& CR2.8 &	\citet{Ebeletal2024}	& PTS	& 0.636$\pm$0.014	& 0.449$\pm$0.016	& 1.216$\pm$0.032	& 1256\\
\hline
Nova 002	&CV3 &	\citet{Ebeletal2016}	& PTS	& 0.500$\pm$0.037	& 0.335$\pm$0.025	& 1.089$\pm$0.086	& 153\\
Allende &	CV$_{\rm oxA}$3.2	& \citet{Ebeletal2016} &	PTS	&0.581$\pm$0.006	&0.571$\pm$0.008	&1.014$\pm$0.019	&4220\\
Allende &	CV$_{\rm oxA}$3.2	& \citet{Simonetal2018} & PTS	&0.310$\pm$0.008	&0.165$\pm$0.004	&0.966$\pm$0.042	&2339\\
Bali	&CV$_{\rm oxB}$3.2	&\citet{Ebeletal2016}	& PTS	&0.746$\pm$0.061	&0.680$\pm$0.041	&1.570$\pm$0.174	&120\\
Leoville	&CV$_{\rm red}$3.1 &\citet{Ebeletal2016}	&PTS	 &0.641$\pm$0.037	&0.519$\pm$0.036	&1.209$\pm$0.082	&186\\
Mokoia	&CV$_{\rm oxB}$3.2	& \citet{Ebeletal2016} & PTS & 0.421$\pm$0.028 &0.405$\pm$0.022	&0.806$\pm$0.066	& 141\\
Tibooburra	&CV$_{\rm oxA}$3	&\citet{Ebeletal2016}	&PTS & 0.254$\pm$0.009	&0.183$\pm$0.005	&0.691$\pm$0.042 &999\\
Vigarano	&CV$_{\rm red}$3.1	&\citet{Ebeletal2016}	&PTS & 0.237$\pm$0.017	&0.143$\pm$0.006	&0.870$\pm$0.117	&406\\
\hline
QUE 94594	&EL3	&\citet{Friedrichetal2022}	&CT	&0.568$\pm$0.041	&0.566$\pm$0.017	&1.032$\pm$0.164	&298\\
\textit{various}	&EL3	&\citet{Schneideretal2002}	&PTS	 &0.461$\pm$0.024	&0.511$\pm$0.021	&0.830$\pm$0.072	&199\\
\hline
\textit{various}	&EH3 &\citet{Schneideretal2002}	&PTS	&0.271$\pm$0.018	&0.269$\pm$0.015	&0.518$\pm$0.049	&135\\
\textit{various}	&EH3	 &\citet{RubinGrossman1987}	&\textit{PTS}	&0.268$\pm$0.008	&0.287$\pm$0.007	&0.499$\pm$0.044	&688\\
\hline
Sharps	&H3.4	&\citet{Friedrichetal2022}	&CT	&0.265$\pm$0.008	&0.284$\pm$0.005	&0.416$\pm$0.022	&665\\
Allegan	&H4	&\citet{MartinMills1978} &\textit{D} &0.631$\pm$0.014	&0.660$\pm$0.009	&0.811$\pm$0.031	&1256\\
Hammond Downs	&H4	&\citet{Kuebleretal1999}	&\textit{PTS}	& 0.435$\pm$0.015	&0.540$\pm$0.017	&0.663$\pm$0.059	&261\\
NWA 2465	&H4	&\citet{Metzler2018}	&D	&0.537$\pm$0.022	&0.494$\pm$0.011	&0.974$\pm$0.059	&761\\
NWA 2465	&H4	&\citet{Metzler2018}	&PTS	&0.448$\pm$0.009	&0.552$\pm$0.010	&0.712$\pm$0.034	&807\\
\hline
Saratov	&L4	&\citet{Friedrichetal2022}	&CT	&0.527$\pm$0.019	&0.522$\pm$0.012	&0.876$\pm$0.049	&550\\
Saratov	&L4	&\citet{Metzler2018}	&PTS	&0.502$\pm$0.011	&0.644$\pm$0.012	&0.740$\pm$0.025	&568\\
Saratov	&L4	&\citet{Metzler2018}	&D	&0.608$\pm$0.017	&0.609$\pm$0.011	&0.997$\pm$0.047	&842\\
\hline
Bjurb\"ole & L/LL4  &\citet{Kuebleretal1999} & \textit{PTS} & 0.570$\pm$0.022 & 0.681$\pm$0.027 & 0.878$\pm$0.083 & 210\\
\hline
ALH 83010	&LL3.3	&\citet{Friedrichetal2022}	&CT	&0.564$\pm$0.022	&0.634$\pm$0.020	&0.816$\pm$0.044	&140\\
Chainpur	&LL3.2	&\citet{Friedrichetal2022}	&CT	&0.697$\pm$0.030	&0.744$\pm$0.019	&1.105$\pm$0.081	&295\\
Kelly	&LL4	&\citet{Kuebleretal1999}	&\textit{PTS}	&0.648$\pm$0.024	&0.788$\pm$0.027	&0.998$\pm$0.081	&222\\
NWA 7545	&LL4	&\citet{Metzler2018}	&D	&0.812$\pm$0.028	&0.829$\pm$0.019	&1.299$\pm$0.063	&458\\
NWA 7545	&LL4	&\citet{Metzler2018}	&PTS	&0.691$\pm$0.016	&0.869$\pm$0.015	&1.088$\pm$0.053	&661\\
Ragland	&LL3.4	&\citet{Friedrichetal2022}	&CT	&0.564$\pm$0.032	&0.571$\pm$0.019	&0.932$\pm$0.089	&229\\
Semarkona	&LL3.0	&\citet{Ebeletal2024}	&PTS	&0.490$\pm$0.011	&0.562$\pm$0.012	&0.786$\pm$0.027	&781\\
\textit{various}	&LL3	&\citet{NelsonRubin2002}	&PTS	&0.660$\pm$0.015	&0.770$\pm$0.017	&1.056$\pm$0.056	&719\\
\hline
NWA 5717 & O3.0-an & \citet{Simonetal2018} & PTS & 0.298$\pm$0.002 & 0.275$\pm$0.002 & 0.620$\pm$0.011 & 12327\\
\hline
\end{tabular}}
\end{table}

\subsection{Sphere-weighted average}

  We coin here the \textit{sphere-weighted average}, that is, the size average weighted by chondrule 3D spherical area: 
\begin{equation}
\label{S/V}
\langle  a_{3D}\rangle_{a_{3D}^2} = 
\frac{\langle  a_{3D}^3\rangle}{\langle  a_{3D}^2\rangle}=\frac{3\pi}{8}\frac{\langle  a_{2D}^2\rangle}{\langle  a_{2D}\rangle}
\end{equation}
where the final equality uses equations \ref{a2D-3D^q weighted} for $q=-1$. This corresponds to the size of a fictive monodisperse chondrule population with the same (total volume)/(total surface) ratio as the actual population\footnote{The interpretation of the sphere-weighted average as three times the (total volume)/(total surface) ratio strictly holds for spherical chondrules. Otherwise the outside surface of a chondrule is greater than that of the equal-volume sphere (more generally defining $a_{3D}$), though not by more than 10 \% for revolution ellipsoids with axes within a factor of 2 of each other. Hence also the expression "sphere-weighted" rather than "outside area-weighted" as a recognition of this invocation of the equal-volume sphere. \ref{Ellipsoids} also displays how the correction factor is modified if one prefers an outside area-weighted average.}. This is the ratio relevant e.g. when calculating geometric opacities \citep[e.g.][]{Marrocchietal2019AOA}, condensation kinetics \citep[e.g.][]{Jacquetetal2026} and many physical problems \citep{Rubin2023}.


   In a thin section or polished block, one reproducible procedure to accurately evaluate this average without making an exhaustive measurement of all chondrules would be the following:
\begin{itemize}
\item[(i)] Select points by making constant steps (e.g. 1 mm, or anything larger than typical chondrules) in the sections (similar to \citet{Doddetal1967,JacquetDoisneau2024}).  
\item[(ii)] Each time the point is inside a chondrule, measure its 2D size (i.e. that of the equal-area disk). Else, move on.
\item[(iii)] Calculate the (unweighted) average of the inverse of all evaluated sizes. This corresponds to $\langle a_{2D}^{-1}\rangle_{a_{2D}^2}$ (in expectation value). Then one can obtain
\begin{equation}
\langle  a_{3D}\rangle_{a_{3D}^2}=\frac{3\pi}{8 \langle a_{2D}^{-1}\rangle_{a_{2D}^2}}
\end{equation}
\end{itemize}

  Such a procedure would also allow to retrieve arithmetic averages. Indeed:
\begin{equation}
\langle a_{2D}\rangle = \frac{\langle a_{2D}^{-1}\rangle_{a_{2D}^2}}{\langle a_{2D}^{-2}\rangle_{a_{2D}^2}}
\end{equation}
And (using equation \ref{a3D}):
\begin{equation}
\langle a_{3D}\rangle = \frac{\pi\langle a_{2D}^{-2}\rangle_{a_{2D}^2}}{2\langle a_{2D}^{-3}\rangle_{a_{2D}^2}}
\end{equation}

   If an exhaustive survey of all chondrules in a section has been made, knowledge of its standard deviation suffices to correct the arithmetic average of $a_{2D}$ to its sphere-weighted counterpart. Indeed:
\begin{equation}
\frac{\langle a_{3D}\rangle_{a_{3D}^2}}{\langle a_{2D}\rangle}=\frac{3\pi}{8}\left(1+RSD(a_{2D})^2\right).
\end{equation}

\begin{figure}
\centering
\includegraphics[width=\textwidth]{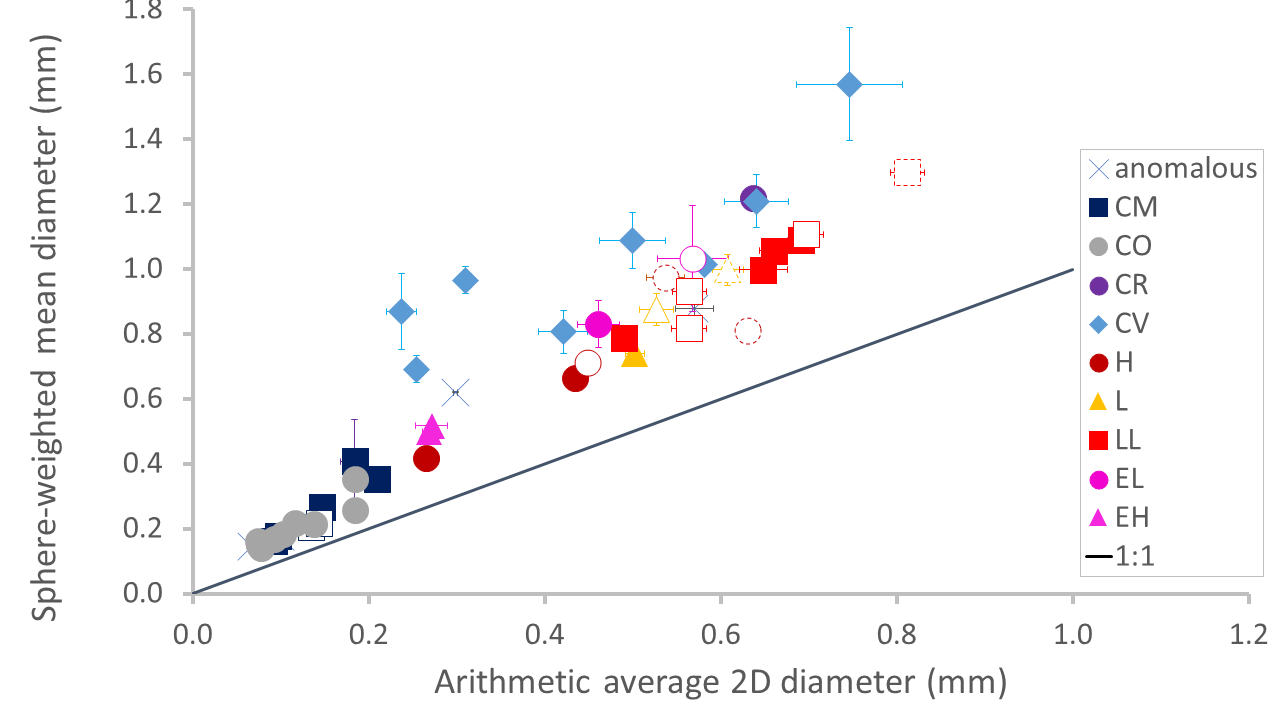}
\caption{Comparison of sphere-weighted average diameter and arithmetic 2D average diameter. Same data sources and presentation as Fig. \ref{2D vs 3D}.}\label{sphere vs 2D}
\end{figure}

\begin{figure}
\centering
\includegraphics[width=\textwidth]{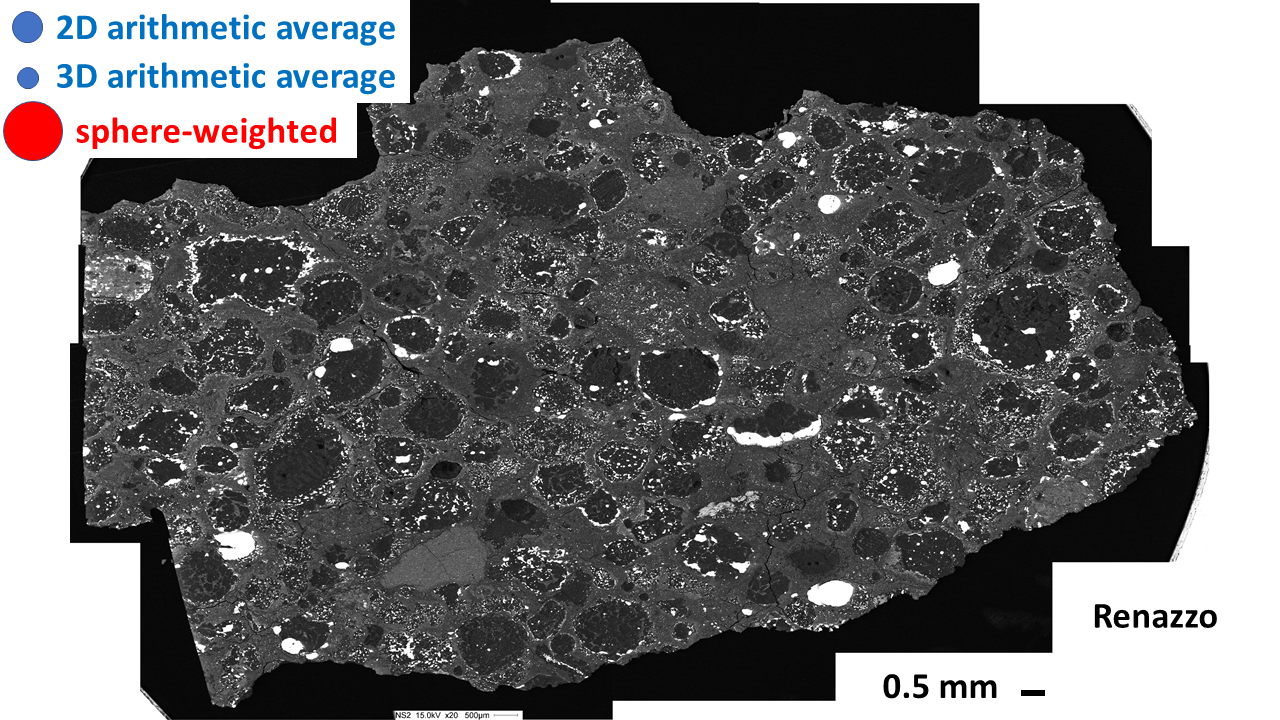}
\caption{Illustration of the different size averages for Renazzo (after the data of \citet{Ebeletal2024}) on a back-scattered electron image of polished section 719sp3 of the Mus\'{e}um national d'Histoire naturelle collection \citep{Jacquetetal2012CC}.}\label{Renazzo}
\end{figure}

\begin{figure}
\centering
\includegraphics[width=\textwidth]{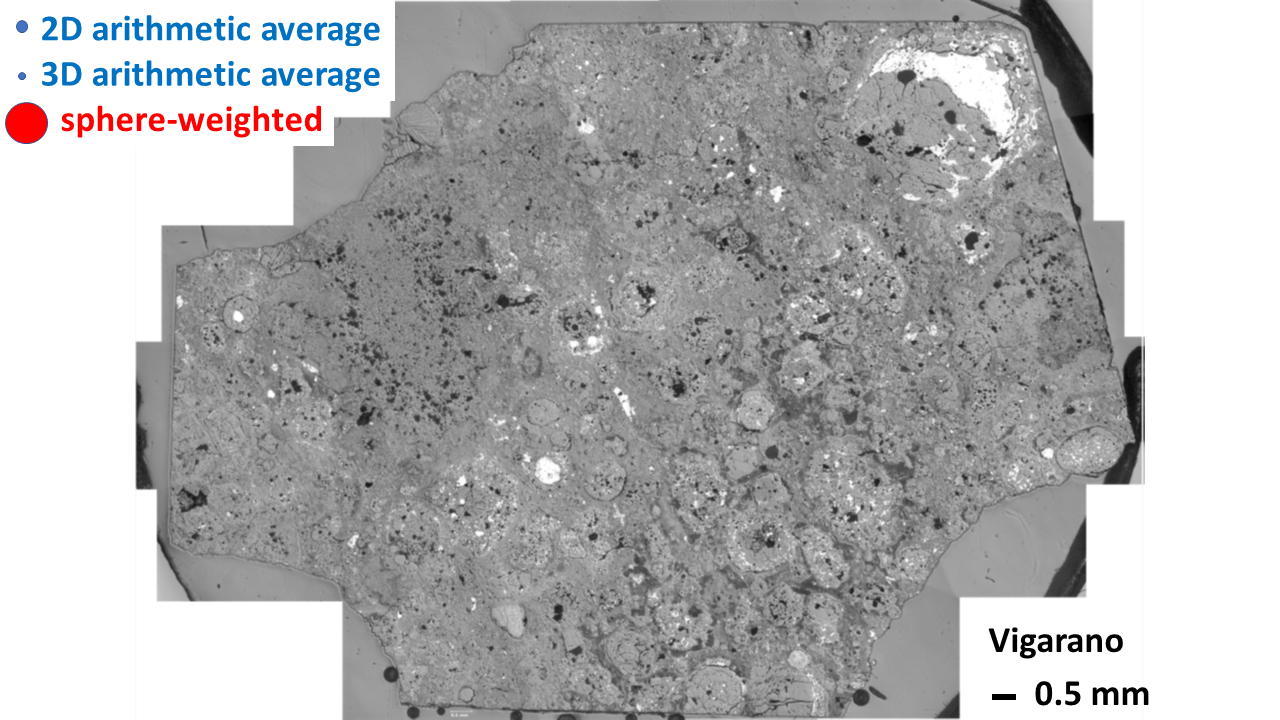}
\caption{Same as Fig. \ref{Renazzo} for a reflected light image of Vigarano section 2793sp2 \citep{Jacquetetal2012CC}. Size data from \citet{Ebeletal2016}.}\label{Vigarano}
\end{figure}

\begin{figure}
\centering
\includegraphics[width=\textwidth]{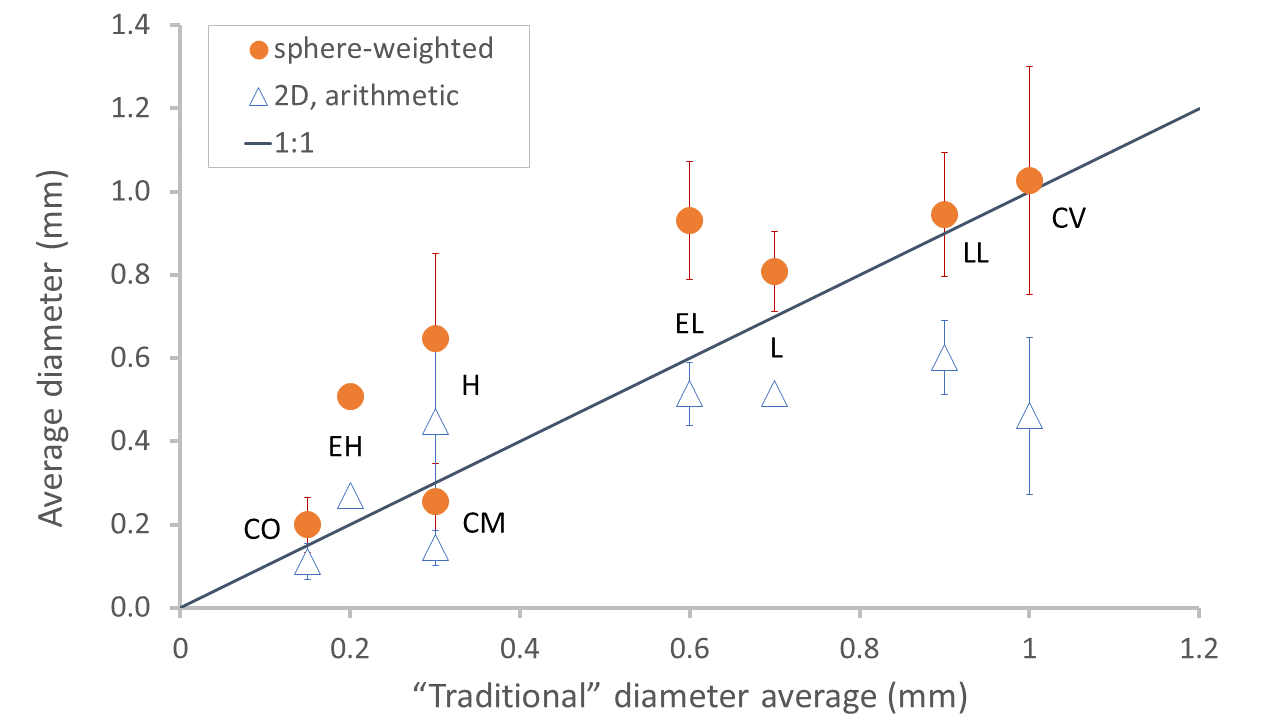}
\caption{Comparison of group averages of size averages with "traditional" size estimates taken from \citet{BrearleyJones1998}. Error bars represent full standard deviation of the averaged dataset, as a measure of their reproducibility. Disaggregation studies were excluded from the averages, following \ref{Inversion}.}\label{size vs tradition}
\end{figure}

  Thus the sphere-weighted average is larger than both 2D and 3D arithmetic averages (Fig. \ref{sphere vs 2D}), owing to the greater weight conferred to larger objects, so is not an estimate thereof. Yet we submit that it better captures what the unprejudiced observer would spontaneously deem a typical size (Fig. \ref{Renazzo}, \ref{Vigarano}). At any rate, these are closer to "traditional" averages reported in the past century (Fig. \ref{size vs tradition}).

\section{Conclusion}
  This work has shown that 3D chondrule size averages can be readily expressed as a function of 2D moments (and vice-versa), and that even size distributions were relatively straightforward to invert. The proof can be even used to assess the effects of deviation to sphericity, provided chondrule sizes are defined by that of the equal-area disk (in 2D) and spheres (in 3D), a general trend in the existing literature that must be encouraged. Thus, 2D data remain competitive tools to assess 3D distributions, within the accuracy limits due to chondrule's departure from sphericity (a few percent underestimates). 2D data are more reliable than disaggregation data which may be affected by differential resistance to fragmentation of larger chondrules, though of little consequence for arithmetic size averages. They can access smaller objects than computed tomography \citep[e.g.][]{Floydetal2024}, which is however steadily improving. In fact, the "missing small chondrules" essentially affect arithmetic size averages, and I suggest that the sphere-weighted average defined herein is  more robust and physically meaningful, in relating to the overall surface/volume ratio.

\section*{Acknowledgments}
This work was supported by ANR PERSEID (ANR-25-CE49-3880). I thank Dr. Dominik Hezel and an anonymous reviewer for their comments leading to an improvement of the comparison with numerical results and the justification of the basic theorem. This paper is dedicated to the memory of the author's \textit{beau-cousin} Etienne de Naurois.

\appendix
\section{Deviation from sphericity}
\label{Ellipsoids}

  We generalize the treatment to an ellipsoidal chondrule of semi-axes $a$, $b$, $c$, whose surface obeys the equation:
\begin{equation}
\left(\frac{x}{a}\right)^2+\left(\frac{y}{b}\right)^2+\left(\frac{z}{c}\right)^2=1
\end{equation}
in a Cartesian orthonormal coordinate system $(x,y,z)$. We define $a_{3D}=\left(abc\right)^{1/3}$ (that is, the radius of the equal-volume sphere). Likewise,  $a_{2D}$ is the radius of the equal-area disk, i.e. $a_{2D}=\left(A(z,\mathbf{n})/\pi\right)^{1/2}$ where $A(z,\mathbf{n})$ represents the area of the (elliptical) intersection between the ellipsoid and the plane orthogonal to (unit) vector $\mathbf{n}$ crossing the $c$ axis at altitude $z$. With these definitions, equation \ref{f(a2D)} still applies, but we need to evaluate anew $\langle f(a_{2D})\rangle_{a_{2D}^2;a_{3D}}$ for any function $f$. We first restrict the averaging to a fixed multiplet $(a,b,c)$ of semi-axes and a fixed orientation of the normal vector $\mathbf{n}$ relative to the axes of the ellipsoid, that is:
 \begin{equation}
\langle f(a_{2D})\rangle_{a_{2D}^2; (a,b,c), \mathbf{n}}\equiv\int_{-\infty}^{+\infty}f\left(\left(\frac{A(z,\mathbf{n})}{\pi}\right)^{1/2}\right)A(z,\mathbf{n})\mathrm{d}z \left(\int_{-\infty}^{+\infty}A(z,\mathbf{n})\mathrm{d}z\right)^{-1}
\end{equation} 

\begin{figure}
\centering
\includegraphics[width=\textwidth]{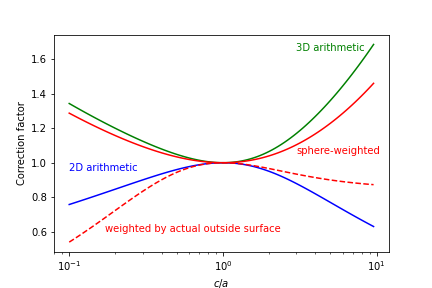}
\caption{Correction factors for our formulas expressing the averages indicated (3D as a function of 2D moments or vice-versa) for revolution ellipsoids of constant $c/a$. 
We include a variant of the sphere-weighted average redefined as weighted by the actual outside surface.}\label{corrections}
\end{figure}

  We consider the volume-conserving transformation 
\begin{equation}
\phi : (x,y,z) \longmapsto a_{3D}\left(\frac{x}{a}, \frac{y}{b}, \frac{z}{c}\right),
\end{equation}
which maps the ellipsoidal surface to the equal-volume sphere. Since $\phi^{-1}(\mathbf{n})\cdot \phi(\mathbf{r})=\mathbf{n}\cdot\mathbf{r}$, a plane orthogonal to $\mathbf{n}$ crossing the $c$ axis at $z$ transforms into a plane crossing the $c$ axis at $a_{3D}z/c$ orthogonal to $\phi^{-1}(\mathbf{n})$. The area $A'(a_{3D}z/c,\phi^{-1}(\mathbf{n}))$ of its intersection with the unit sphere may be found by writing the conservation of the transformed volume between the intersection ellipses at $z$ and $z+\mathrm{d}z$:
\begin{equation}
A(z,\mathbf{n})\mathrm{d}z\mathrm{cos}(\mathbf{c},\mathbf{n})=A'\left(\frac{a_{3D}z}{c},\phi^{-1}(\mathbf{n})\right)\mathrm{d}\left(\frac{a_{3D}z}{c}\right)\mathrm{cos}(\mathbf{c},\phi^{-1}(\mathbf{n}))
\end{equation}   
where $\mathbf{c}$ is the vector of length $c$ directed by the corresponding axis of the ellipsoid. Defining similarly $\mathbf{a}$ and $\mathbf{b}$, we have:
\begin{equation}
\frac{A(z,\mathbf{n})}{A'(a_{3D}z/c,\phi^{-1}(\mathbf{n}))}=\frac{a_{3D}}{\sqrt{\left(\mathbf{a}\cdot\mathbf{n}\right)^2+\left(\mathbf{b}\cdot\mathbf{n}\right)^2+\left(\mathbf{c}\cdot\mathbf{n}\right)^2}}\equiv \alpha_{\rm corr}^2
\end{equation}
and thus
 \begin{eqnarray}
\langle f(a_{2D})\rangle_{a_{2D}^2;(a,bc),\mathbf{n}}&=& \int_{-\infty}^{+\infty} f\left(\alpha_{\rm corr}\left(\frac{A'(z',\phi^{-1}(\mathbf{n}))}{\pi}\right)^{1/2}\right) A'(z',\phi^{-1}(\mathbf{n})) \mathrm{d}z' \nonumber\\
\left(\int_{-\infty}^{+\infty}A'(z',\phi^{-1}(\mathbf{n}))\mathrm{d}z'\right)^{-1}\nonumber\\
&=& \langle f(\alpha_{\rm corr}a_{2D})\rangle_{a_{2D}^2;(a_{3D},a_{3D},a_{3D})}\nonumber\\
&=& \langle f(a_{2D})\rangle_{a_{2D}^2;(\alpha_{\rm corr}a_{3D},\alpha_{\rm corr}a_{3D},\alpha_{\rm corr}a_{3D})}
\end{eqnarray} 
where the two final forms refer to results for spheres
.
  In particular, for a constant exponent $q$:
\begin{equation}
\langle a_{2D}^q\rangle_{a_{2D}^2;(a,b,c), \mathbf{n}}=\alpha_{\rm corr}^q a_{3D}^q\frac{3W_{q+3}}{2}
\end{equation}
Averaging over all relative orientations of $\mathbf{n}$ yields the following correction factor relative to the spherical case
:
\begin{equation}
\alpha_{\rm corr,q}=\langle \alpha_{\rm corr}^q\rangle_{\mathbf{n};(a,b,c)}=\left(\frac{c}{a}\right)^{q/6}\int_0^{1}\frac{\mathrm{d}u}{\left(1+\left(\left(\frac{c}{a}\right)^2-1\right)u^2\right)^{q/4}}
\end{equation}
where the final equality assumes an isotropic distribution (not appropriate e.g. in case of a postaccretional foliation) and $a=b$ (i.e. a revolution ellipsoid). The resulting correction factors on the 2D-3D conversion formulas are plotted in Fig. \ref{corrections} for a constant shape. Axes within a factor of two of each other (as for most chondrules, e.g. \citet{Friedrichetal2022,Charlesetal2018}) should not incur more than 6 \% correction. This correction factor is subject to an average over all possible shapes (here $c/a$ ratios). So 3D size averages estimated from 2D sizes are underestimates (except if an actual surface-weighted average is adopted, where we would have an overestimate), but only by a few percent.

\section{Explicit PDF inversion from 2D to 3D}
\label{Inversion}

  In this appendix, we derive explicit relationships between the cumulative functions (by chondrule number) of $a_{3D}$ and $a_{2D}$, named here $n_{3D}$ and $n_{2D}$ respectively. 

  We note that for any function $g$, dividing equation \ref{f(a2D)} (with the second form of equation \ref{f(a2D) at fixed a3D} injected) with $f:a\rightarrow a^2g(a)$ by the same one with $f:a\rightarrow a^{-2}$ yields:
\begin{eqnarray}
\label {g(a2D)}
\langle g(a_{2D})\rangle 
&=& \langle a_{3D}\int_0^{1}g(a_{3D}\sqrt{1-Z^2})\mathrm{d}Z\rangle/\langle a_{3D}\rangle
\end{eqnarray}

\subsection{Forward problem}

  We first solve anew the "forward" problem (from 3D to 2D; \citet{CuzziOlson2017}) by applying 
equation \ref{g(a2D)} to function 
\begin{equation}
g:a_{2D}\longrightarrow H_0(a_{2D}-x)
\end{equation}
with $x$ some constant threshold size and $H_0$ the Heaviside function returning 1 for a strictly positive argument and zero otherwise. Since 
\begin{equation}
\int_0^1 H_0 \left(a_{3D}(1-Z^2)^{1/2}-x\right)\mathrm{d}Z=H_0(a_{3D}-x)\sqrt{1-\left(\frac{x}{a_{3D}}\right)^2},
\end{equation}
we have:
\begin{equation}
1-n_{2D}(x)=\langle H_0(a_{2D}-x)\rangle = \langle H_0(a_{3D}-x)\left(a_{3D}^2-x^2\right)^{1/2}\rangle/\langle a_{3D}\rangle
\end{equation}
Hence:
\begin{equation}
\label{n2D forward}
n_{2D}(x)=1-\frac{1}{\langle a_{3D}\rangle}\int_x^{+\infty}\frac{\mathrm{d}n_{3D}}{\mathrm{d}a_{3D}}\sqrt{a_{3D}^2-x^2}\mathrm{d}a_{3D}
\end{equation}
We hereby find that the probability of a 2D section being \textit{larger} than $x$ (i.e. $1-n_{2D}(x)$) is equal to the integral of the 3D size density probability of the \textit{sectioned} chondrules (i.e. $(a_{3D}/\langle a_{3D}\rangle)dn_{3D})/da_{3D}$, as the probability of sectioning a given chondrule is proportional to $a_{3D}$) \textit{mutliplied} by the probability of the section being larger than $x$ (i.e. $\sqrt{a_{3D}^2-x^2}/a_{3D}$). This is how the forward problem has been hitherto calculated in the literature \citep{Eisenhour1996,CuzziOlson2017}, and thus brings additional confidence in the formalism.
Equation \ref{n2D forward} can be differentiated as:
\begin{equation}
\frac{\mathrm{d}n_{2D}}{\mathrm{d}a_{2D}}=\frac{a_{2D}}{\langle a_{3D}\rangle}\int_{a_{2D}}^{+\infty}\frac{\mathrm{d}n_{3D}}{\mathrm{d}a_{3D}}\frac{\mathrm{d}a_{3D}}{\sqrt{a_{3D}^2-a_{2D}^2}}
\end{equation}
It is in fact possible to derive anew equation \ref{f(a2D) original} (if only for the spherical chondrule case) from this equation, that is from the "traditional" reasoning recounted in the previous paragraph, without reference to our opening argument in the main text. Indeed:
\begin{eqnarray}
\label{f(a2D) defined}
\langle f(a_{2D})\rangle_{\pi a_{2D}^2} &\equiv& \int_0^\infty f(a_{2D})\frac{\mathrm{d}n_{2D}}{\mathrm{d}a_{2D}}\pi a_{2D}^2\mathrm{d}a_{2D}\left(\int_0^\infty \frac{\mathrm{d}n_{2D}}{\mathrm{d}a_{2D}}\pi a_{2D}^2\mathrm{d}a_{2D}\right)^{-1}\nonumber\\
\end{eqnarray}
Injecting the previous expression in the numerator of the above 
 yields:
\begin{eqnarray}
\int_0^{+\infty} f(a_{2D})\frac{\mathrm{d}n_{2D}}{\mathrm{d}a_{2D}}\pi a_{2D}^2\mathrm{d}a_{2D}&=&\frac{1}{\langle a_{3D}\rangle}\int_0^{+\infty}\mathrm{d}a_{3D}\int_0^{a_{3D}}\frac{\mathrm{d}n_{3D}}{\mathrm{d}a_{3D}}\pi a_{2D}^2f(a_{2D})\frac{a_{2D}\mathrm{d}a_{2D}}{\sqrt{a_{3D}^2-a_{2D}^2}}\nonumber\\
&=& \frac{1}{\langle a_{3D}\rangle}\int_0^{+\infty}\mathrm{d}a_{3D}\frac{\mathrm{d}n_{3D}}{\mathrm{d}a_{3D}}\int_0^{a_{3D}}\pi a_{2D}^2f(a_{2D})\mathrm{d}z\nonumber\\
&=& \frac{1}{2\langle a_{3D}\rangle}\int_0^{+\infty} \frac{4\pi}{3}a_{3D}^3\langle f(a_{2D})\rangle_{\pi a_{2D}^2;a_{3D}}\frac{\mathrm{d}n_{3D}}{\mathrm{d}a_{3D}}\mathrm{d}a_{3D}
\end{eqnarray}
where the second expression has made use of Fubini's theorem, before a change of variable  of the second integration from $a_{2D}$ to $z=\sqrt{a_{3D}^2-a_{2D}^2}$. The same manipulations apply to the denominator of 
equation \ref{f(a2D) defined}, which only differs by the substitution of $f(a_{2D})$ by 1, hence the final result.

\subsection{Inverse problem}

  For the inversion problem (from 2D to 3D), we actually do not use the above solution, but apply again the first form of equation \ref{g(a2D)} to a new function
\begin{equation}
g:a_{2D}\longrightarrow\frac{H_0(a_{2D}-x)}{\sqrt{a_{2D}^2-x^2}}
\end{equation}
with $x$ again some constant threshold. 
Since\footnote{With the integrand nonzero only for $Z<Z_{\rm max}=\sqrt{1-(x/a_{3D})^2}$, we can restrict the interval of integration to $[0,Z_{\rm max}]$ and change there the variable of integration to $\alpha$ such that $Z=Z_{\rm max}\mathrm{cos}\alpha$.}
\begin{equation}
\int_0^1 \frac{H_0(a_{3D}(1-Z^2)^{1/2}-x)}{\sqrt{a_{3D}^2(1-Z^2)-x^2}}\mathrm{d}Z=\frac{\pi}{2a_{3D}}H_0(a_{3D}-x)
\end{equation}
we have
\begin{equation}
\langle H_0(a_{2D}-x)\left(a_{2D}^2-x^2\right)^{-1/2}\rangle=\frac{\pi}{2\langle a_{3D}\rangle}\langle H_0(a_{3D}-x)\rangle=\langle a_{2D}^{-1}\rangle \left(1-n_{3D}(x)\right)
\end{equation}
where the final equality has made use of equation \ref{a3D}. Hence:
\begin{eqnarray}
\label{n3D}
n_{3D}(x)&=& 1-\langle \frac{H_0(a_{2D}-x)}{\sqrt{a_{2D}^2-x^2}}\rangle \langle a_{2D}^{-1}\rangle^{-1}\nonumber\\
&=& 1-\frac{1}{\langle a_{2D}^{-1}\rangle}\int_x^{+\infty}\frac{\mathrm{d}n_{2D}}{\mathrm{d}a_{2D}}\frac{\mathrm{d}a_{2D}}{\sqrt{a_{2D}^2-x^2}}
\end{eqnarray}

  Therefore, if $n_{2D}$ is differentiable twice, after an integration per part:
\begin{equation}
\frac{\mathrm{d}n_{3D}}{\mathrm{d}a_{3D}}=\frac{a_{3D}}{\langle a_{2D}^{-1}\rangle}\int_{a_{3D}}^{+\infty}\frac{\mathrm{d}}{\mathrm{d}a_{2D}}\left(\frac{1}{a_{2D}}\frac{\mathrm{d}n_{2D}}{\mathrm{d}a_{2D}}\right)\frac{\mathrm{d}a_{2D}}{\sqrt{a_{2D}^2-a_{3D}^2}}
\end{equation}

\begin{figure}
\centering
\includegraphics[width=\textwidth]{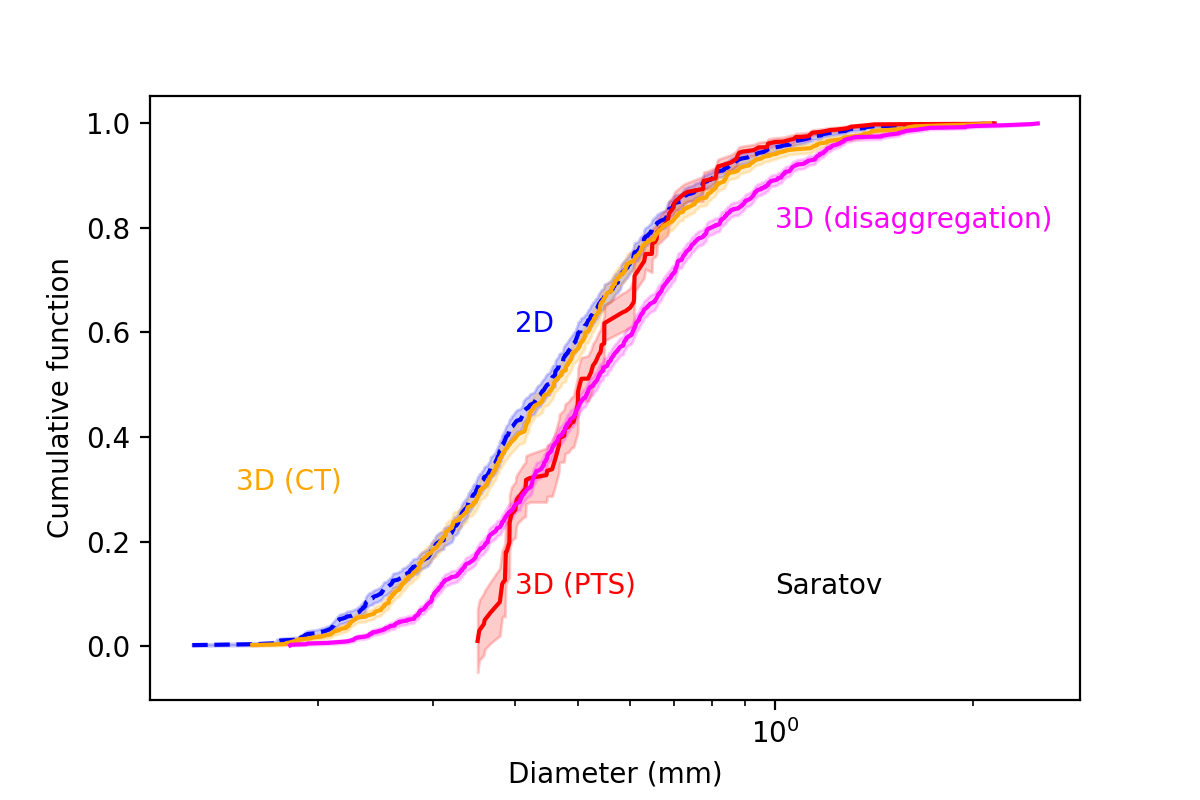}
\caption{Cumulative functions for Saratov chondrules from polished thin section \citep["2D";][]{Metzler2018}, its 3D inversion following this appendix ("3D (PTS)"), disaggregation data of \citet{Metzler2018} and computed tomography \citep["3D (CT)";][]{Friedrichetal2022}. Shaded areas represent statistical error bars ($\pm 1\sigma$) following \ref{Error}. Disaggregation data are biased toward larger chondrules. The 3D CT data largely overlap with the \textit{2D} ones of \citet{Metzler2018}, suggesting sample heterogeneity between the two studies.}\label{Saratov}
\end{figure}

\begin{figure}
\centering
\includegraphics[width=\textwidth]{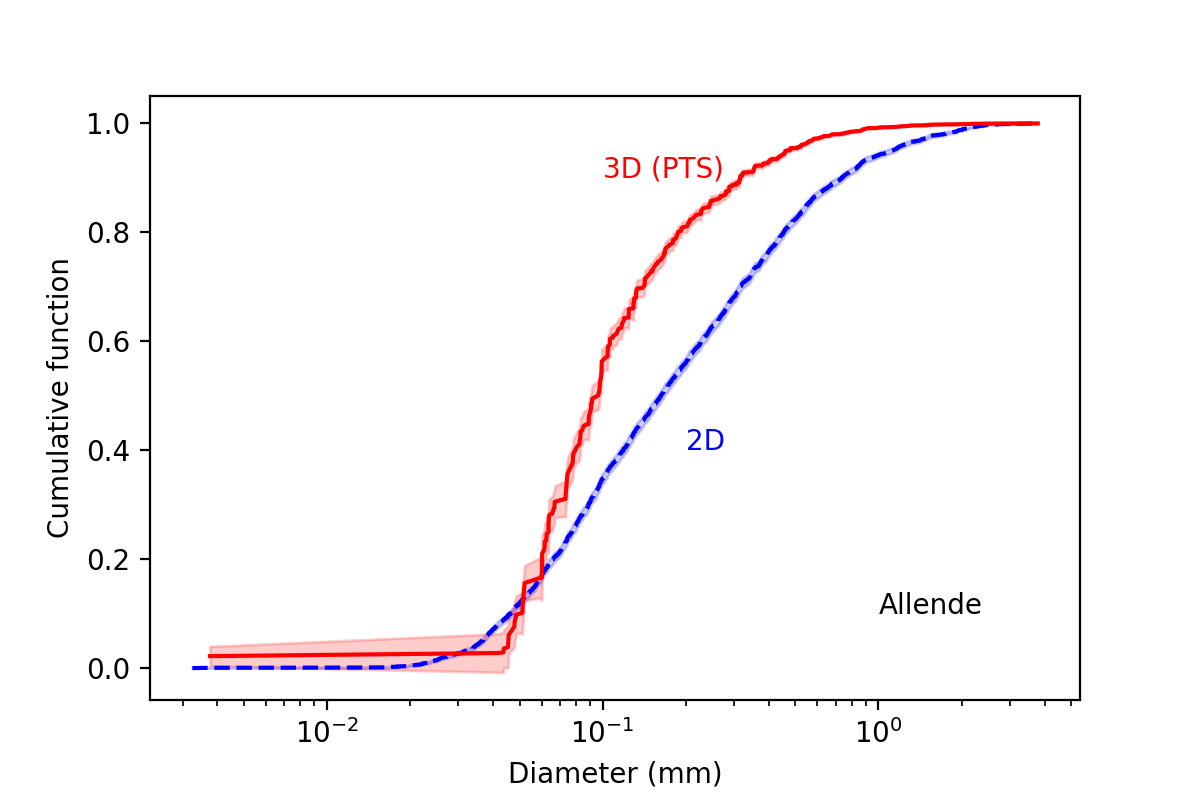}
\caption{Cumulative functions for Allende chondrules from polished thick section \citep["2D";][]{Simonetal2018} and its 3D inversion following this appendix ("3D (PTS)").}\label{Allende}
\end{figure}

  For a given set of $n$ measured $a_{2D,i}$, assumed to be sorted in increasing order (from $i=1$ to $i=n$), the first form of equation \ref{n3D} can be estimated as:
\begin{equation}
n_{3D}(x)_{\rm est}=1-\sum_{j=1}^{n}\frac{H_0(a_{2D,j}-x)}{\sqrt{a_{2D,j}^2-x^2}}\left(\sum_{j=1}^{n}\frac{1}{a_{2D,j}}\right)^{-1}
\end{equation}
However, this estimator decreases between two consecutive values of measured $a_{2D,i}$ (and even diverges to $-\infty$ at each $a_{2D,i}^-$), even though $n_{3D}$ increases by definition. It can thus be useful only at each $a_{2D,i}$ (which corresponds to the limit for $x\rightarrow a_{2D,i}$ with $x>a_{2D,i}$), as:
\begin{equation}
n_{3D}(a_{2D,i})_{\rm est}=1-\sum_{j=i+1}^{n}\frac{1}{\sqrt{a_{2D,j}^2-a_{2D,i}^2}}\left(\sum_{j=1}^{n}\frac{1}{a_{2D,j}}\right)^{-1}
\end{equation}
 Remaining apparent deviations to monotonicity (or even positivity), due to too close consecutive values of $a_{2D,i}$ should be ignored since they are less useful estimators than any increasing interpolation between surrounding estimates. Fig. \ref{Saratov} and \ref{Allende} illustrate two such PDF inversions. The steeper slopes (that is, more peaked PDF) show that, in general, the various sectioning geometries broaden the original (3D) distribution. 
 At low sizes, the reconstructed cumulative functions are not resolvable from zero because the smallest 2D chondrule sections are compatible with a "tail" due to the nonequatorial sectioning of bigger chondrules.

  Disaggregation studies may be somewhat biased in favor of large chondrules (Fig. \ref{Saratov}). Indeed, the chondrules separated by \citet{Metzler2018} represent about 0.4 g (NWA 2365, H4), 0.7 g (Saratov, L4) and 0.4 g (NWA 7545, LL4), despite "a few grams" being crushed for each. Thus most chondrules are fragmented. This leaves room for differential resistance to fragmentation in favor of bigger objects, which would be then overrepresented. Although this effect is not appreciable for arithmetic size averages (Fig. \ref{3D vs litt}; \citet{Metzler2018}) it appears significant for the sphere-weighted ones (19-37 \% higher than the thin section-based estimates from the \citet{Metzler2018} datasets, beyond the few percents systematic underestimates expected from deviation to sphericity in \ref{Ellipsoids}). These are indeed more sensitive to larger chondrules.

\section{Error estimate}
\label{Error}

  This appendix estimates the ratio of the expectation values of two random variables $X$ and $Y$ (in this paper, different functions of chondrule sizes) using $n$ independent realizations $(X_i, Y_i)$ thereof (that is, $n$ different chondrules). Our estimator is:
\begin{eqnarray}
\left(\frac{\langle X\rangle}{\langle Y\rangle}\right)_{\rm est} & \equiv & \sum_{i=1}^{n}X_i \left( \sum_{i=1}^{n}Y_i\right)^{-1}\nonumber\\
& = &  \frac{\langle X\rangle}{\langle Y\rangle}\left(1+\frac{1}{n}\sum_{i=0}^{n}\delta \tilde{X}_i\right)\left(1+\frac{1}{n}\sum_{i=0}^{n}\delta \tilde{Y}_i\right)^{-1}
\end{eqnarray}
with $\delta\tilde{X}_i\equiv X_i/\langle X\rangle -1$ (and similarly for $\delta\tilde{Y}_i$).
  
  To lowest order as $n\rightarrow +\infty$:
\begin{equation}
E\left(\left(\frac{\langle X\rangle}{\langle Y\rangle}\right)_{\rm est}\right)=\frac{\langle X\rangle}{\langle Y\rangle}\left(1+\frac{RSD(Y)}{n}\left(RSD(Y)-\mathrm{Cor}(X,Y)RSD(X)\right)+o\left(\frac{1}{n}\right)\right) 
\end{equation}
\begin{equation}
RSD\left(\left(\frac{\langle X\rangle}{\langle Y\rangle}\right)_{\rm est}\right)\sim \left(\frac{RSD(X)^2+RSD(Y)^2-2RSD(X)RSD(Y)\mathrm{Cor}(X,Y)}{n}\right)^{1/2}.
\end{equation}

\bibliographystyle{elsarticle-harv} 
\bibliography{bibliography}







\end{document}